\documentclass[12pt,a4paper]{amsart}
\usepackage[utf8]{inputenc}
\usepackage{amsmath}
\usepackage{amssymb}
\usepackage{tikz}
\usepackage{color}
\usepackage[left=3cm,right=2cm,top=3cm,bottom=2cm]{geometry}\usepackage[square,sort,comma,numbers]{natbib}

\title{Some {applications of Griffiths} theorem \\ {in theory} of Feynman integrals}

\author[V. A. Golubeva]{Valentina A. Golubeva}
\address{Department of Applied Mathematics and Physics\\
Moscow Aviation Institute
(National Research University)\\
4 Volokolamskoe shosse\\
125080 Moscow, Russia}
\email{goloubeva@yahoo.com}

\author[A. N. Ivanov]{Alexey N. Ivanov}
\address{Department of Mathematics\\
National Research University
Higher School of Economics\\
6 Usacheva Street\\
119048 Moscow, Russia}
\email{anivanov$_{-}$1@edu.hse.ru}

\begin{document}

\begin{abstract}
{The present paper provides a method for finding partial differential equations satisfied by the Feynman integrals for diagrams of various types, using the Griffiths theorem on the reduction of poles of rational differential forms. As an application, an algorithm for computing partial differential equations satisfied by Feynman integrals for diagrams of a ladder type is described.}

\end{abstract}

\maketitle

\section{Introduction}

\vspace{5mm}

Feynman integrals studied in the present paper enter the expressions for the elements of the $S$-matrix in scattering problems of quantum field theory. To fix ideas,
consider, for example, an initial state of a quantum system at the moment $t = -\infty$, characterized by  some numbers of fermions and photons in {specified} individual states. The problem consists in describing the state of this system at the moment $t=+\infty$. The answer is given in terms of a wave function $\Phi(t)$, that describes the states of the fields under consideration and {satisfies} the Schr{\"o}dinger equation
\begin{equation} i~ \frac{d\Phi(t)}{dt}=V(t)\Phi(t). \end{equation}

\noindent The general solution of this equation allows one to connect the values of the wave function at $t=-\infty$ and $t=+\infty$ by a transformation of the form
\begin{equation} \Phi(+\infty)=S \Phi(-\infty). \end{equation}

\noindent The operator $S$ entering this expression is unitary and is called the scattering matrix (or the $S$-matrix). The perturbation theory is {the most} widely known method for solving the scattering problem. The asymptotic series for the elements of the $S$-matrix provided by the perturbation theory are sums of Feynman integrals corresponding to the Feynman diagrams with fixed number of internal lines.

The Feynman integrals of the quantum field theory are analytic functions with singularities, depending on certain physical parameters. The singularities occur along hypersurfaces, called singular varieties.
The behavior of a Feynman integral in the neighborhood of its singular variety is similar to the behavior of hypergeometric functions. Analytical properties of Feynman integrals can be studied by various methods, for example, by means of series expansions near the singular variety (see \cite{Kersh}). These series expansions are derived from systems of partial differential equations for these integrals (see \cite{Gol, Gnghm}). In the present paper, we give a method for constructing such a system of partial differential equations for general Feynman integrals, based on the Griffiths theorem \cite{Gr} on the reduction of poles of rational differential forms.

The structure of the paper is as follows. In Sec. 2, {Feynman} diagrams and parametric representations of Feynman {integrals} are defined. In Sec. 3, we introduce the necessary algebraic tools: the Griffiths Theorem and the Macaulay Theorem. In Sec. 4, we apply the Griffiths theorem to the problem of finding systems of partial differential equations for Feynman integrals corresponding to different types of diagrams ({Theorems 1 and 2}).

{\bf Acknowledgements.} The work was supported in part by the Simons Foundation.

\vspace{5mm}

\section{Parametric {representations} of Feynman {integrals}}

\vspace{5mm}

Each {term} of the expansion of {the} $S$-matrix {in a} series corresponds to some Feynman diagram. The Feynman diagram is {a} graph with lines topologically representing the propagation {of} elementary particles and vertices representing {interactions} of the elementary particles \cite{Nak}.

The graph $G$ consists {of} elements of two types, the lines $\{l_i, i \in \Omega\}$ {and} the vertices $\{V_k, k \in \Theta\}$. {There is the} map $i = i_1 \times i_2: \Omega \rightarrow \Theta \times \Theta$ {sending} each line $l_j ~ (j \in \Theta)$ to the pair of vertices $i_1(j), i_2(j)$, called the initial and final vertices of the line respectively. {A} line $l$ is incident {with a} vertex $V$ if it has this vertex as the initial or final vertex. {{A} (minimal) loop of the graph is a sequence {of} lines $l_1, l_2, ..., l_k$ such that $l_i$, $l_{i+1}$ are incident {with some} vertex {$V_{i+1}$} for each $i=1,\ldots,k$, {while $l_1$ and $l_k$ are incident with some vertex $V_1$},  and the vertices $V_1,\ldots,V_k$ are pairwise distinct}. {A} loop can be formed by {just} one line. A path {$\gamma$} in the graph $G$ is a sequence {of} lines $l_{\gamma(1)}, l_{\gamma(2)}, ..., l_{\gamma(r)}$ such that:  {1) neither}  $l_{\gamma(i)}$ is a loop; {2)} $\gamma(i) \neq \gamma(j)$ {whenever} $1 \leq i \neq j \leq r$; {{3)} $l_{\gamma(i)}$, $l_{\gamma(i+1)}$ are incident with a vertex $V_i ~ (1 \leq i \leq r-1)$; 4) $V_i\neq V_{i+1}$ for $i=1,\ldots,r-2$. Denote by $V_0$ the vertex of $l_{\gamma(1)}$, different from $V_1$, and by $V_r$ the vertex of $l_{\gamma(r)}$, different from $V_{r-1}$. If the vertices $V_0, V_1, ..., V_r$ are pairwise distinct, we will say that the path is simple.
 In this case the path $\gamma$ starts at $V_0$ and end at $V_r$.} Denote by $N$ and $n$ the numbers of elements of the sets $\Omega$ and $\Theta$, i. e. $|\Omega|=N, |\Theta|=n$. If $c$ is the number of connected {components of $G$,} then {the number {of} independent loops {of} $G$ {is $h=N-n+c$.}} If {several graphs are considered,} then the {above} sets {and} numbers corresponding to {a} graph $G$ will be denoted by $\Omega_G, \Theta_G$, $N(G), n(G)$.

{A} graph $G$ is called {a} Feynman diagram \cite{SpWe}{,} if {partitions of} $\Theta$ and $\Omega$ {of the following form are given:
\begin{equation} \Theta=\Theta^E \sqcup \Theta^I,  \ \ \Omega=\Omega^M \sqcup \Omega^0. \end{equation}
}
\noindent The vertices $V_j$ for $j \in \Theta^E$ are called external and the vertices $V_j$ for $j \in \Theta^I$ are called internal; the lines $l_j$ for $j \in \Omega^M$ are called massive and the lines $l_j$ for $j \in \Omega^0$ are called massless.

{If $G$} is connected{,} then we denote by $G^{\infty}$ the graph which is obtained {from $G$} by {adding} the infinity vertex $V_{\infty}$ that is connected with all external vertices {by} lines. {Each line of a Feynman diagram is endowed with some orientation,} but the Feynman integral defined below is independent {of} this orientation.

A subgraph $H$ {of $G$} is a graph such that $\Omega_{H} \subset \Omega_{G}, \Theta_{H} \subset \Theta_{G}$ and $i_{H}=i_{G}|_{\Omega_{H}}$. A subgraph $T_r$ of a graph $G$ is called a $r$-tree  if it {contains no} loops and {if} the number of its connected components is equal to $r$. A spanning $r$-tree of $G$ is a $r$-tree which includes {all} the vertices of $G$.

\vspace{2mm} To each line $l_j~ (j \in \Omega)$ {of} $G${, one associates a} parameter $\alpha_j~ (j=1,...,N)$ (these parameters are called the Feynman parameters). For every subset $\eta \subset \Omega${,} we will write $\alpha(\eta)=\prod\limits_{j \in \eta} \alpha_j.$ Define the polynomial $U(\alpha)=\sum\limits_{T_1 \subset G}\alpha(\Omega-T_1)$, where the sum is taken over all spanning trees $T_1$ {of} $G$. Let $\chi \subset \Theta^E$ be {an} arbitrary subset of the set {of} external vertices {of} $G$. Define another polynomial{,} $W_{\chi}(\alpha)=\sideset{}{'}\sum\limits_{T_2 \subset G}^{} \alpha(\Omega-T_2)$, where the sum is taken over all spanning 2-trees $T_2$ {of} $G$ such that its connected components contain subsets $\chi$ and $\Theta^E-\chi$ {respectively}.

Let $G$ be {a connected} graph. Define the space $X_G \subset (\mathbb{C}^{D})^{n^E}$ {of external} momentum vectors {satisfying} the conservation law
\begin{equation} X_G={\left\{ p_k, k \in \Theta^E ~ | ~ \sum\limits_{k \in \Theta^E} p_k = 0 \right\}} . \end{equation}
Assume that a symmetric bilinear form is given on $(\mathbb{C}^{D})^{n^E}$ whose quadratic form is positive definite when restricted to the real subspace $(\mathbb{R}^{D})^{n^E}$ of $(\mathbb{C}^{D})^{n^E}$. {So we can form the invariants
\begin{equation}\label{invar} s(\chi)=\left( \sum\limits_{i \in \chi} p_i \right)^2 \end{equation}
\noindent from the external momentum vectors $p_i ~ (i \in \Theta^E)$,
where $\chi$ is a} non-empty proper subset of the set $\Theta^E$. Due to the conservation law these invariants satisfy the relations
\begin{equation}\label{rel1} s(\chi)=s(\Theta^E - \chi), \end{equation}
\begin{equation}\label{rel2} s(\chi_1 \cup \chi_2 \cup \chi_3) = s(\chi_1 \cup \chi_2) + s(\chi_1 \cup \chi_3) + s(\chi_2 \cup \chi_3) - s(\chi_1) - s(\chi_2) - s(\chi_3),\end{equation}
\noindent where $\chi_1, \chi_2, \chi_3$ are non-empty proper disjoint subsets of $\Theta^{E}$.  If $D \geq n^E - 1$, {then} the relations (\ref{rel1}), (\ref{rel2}) are all the relations {on} the invariants $s(\chi)$ {that are consequences of} their definition (\ref{invar}) (about the relations on the invariants for the case $D < n^E - 1$ see \cite{MiGh}). Hence, we can fix {a collection} of subsets $\mathcal{X}=\{\chi_i ~ | ~ \chi_i \subset \Theta^E \},~ |\mathcal{X}|=\frac{1}{2} n^E (n^E - 1)$, such that all invariants $s(\chi)$ are expressed {in terms of} $s_i=s(\chi_i)${,} and these invariants are {linearly} independent (for example, $\mathcal{X}=\big\{ \{i\}, \{j, k\} ~ | ~ i,j,k \in \Theta^E - {i_0} \big\}$, where  $i_0 \in \Theta^E$ is some fixed external vertex).

So to each graph $G$ we can {associate a} set of complex parameters $s_i${,} which are the invariants defined above. Besides, {to each internal line $l_j ~ (j \in \Omega)$ of $G$, one associates a} complex parameter $z_j$ called the squared mass for that line. Then the parametric representation of the Feynman integral for the given graph $G$ is defined by the following formula{:}
\begin{equation}\label{Feynman} F(s,z)=\int\limits_{\Gamma} \frac{U(\alpha)^{N-\frac{D}{2}(h+1)}}{Q(\alpha, s, z)^{N-\frac{D}{2}h}} ~\omega, \end{equation}

\noindent where
\begin{equation} \omega=\sum\limits_{i=1}^{N} (-1)^i \alpha_i d\alpha_1 \wedge ... \wedge \widehat{d\alpha_i} \wedge ... \wedge d\alpha_N, \end{equation}
\begin{equation}\label{denom} Q(\alpha, s, z) = \frac{1}{2} \sum\limits_{\chi \subset \Theta^E} s(\chi) W_{\chi}(\alpha) - U(\alpha) \sum\limits_{i=1}^{N} \alpha_i z_i{,} \end{equation}

\noindent and $\Gamma$ is {a $(N-1)$-dimensional}  piecewise smooth hypersurface {lying} in the first quadrant of the real subspace $\mathbb{R}^{N}$ of $\mathbb{C}^{N}$ such that its boundary {lies} on the coordinate hyperplanes $\{ \alpha_1=0 \}, ..., \{ \alpha_N=0 \}$. This hypersurface can be chosen {in  such a way} that it {does not} intersect the variety $\{ Q=0 \}$.

\vspace{5mm}

\section{Griffiths theorem on reduction of {a} pole of {a} rational differential form}

\vspace{5mm}

Consider {a rational} differential $n$-form {on the projective space $\mathbb{C}\mathbb{P}^n$
with homogeneous coordinates $\alpha_0,\ldots,\alpha_n$},
\begin{equation}\label{ratl_form}
 \gamma = \frac{R(\alpha)}{Q(\alpha)^k} ~\omega \ \ \ (k \geq 1), \end{equation}

\noindent where $R$ and $Q$ are {homogeneous} polynomials of {degrees} $p$ and $q$
{in $\alpha=(\alpha_0,\ldots,\alpha_n)$,} such that $kq=p+n+1$. The form $\gamma$ is a rational differential form {with pole} of {order} $k$ on the algebraic variety $\{ Q=0 \}$ in $\mathbb{C}\mathbb{P}^n$. Below we will use the following Griffiths theorem on reduction of {a} pole of {a} rational differential form.

\vspace{2mm}\noindent\textbf{Griffiths  theorem \cite{Gr}.~} \textit{Let $\gamma$ be a rational
differential $n$-form {on} the projective space $\mathbb{C}\mathbb{P}^n$ with pole {of order} $k \geq 1$
on the variety $\{ Q(\alpha)=0 \}${, given by formula \eqref{ratl_form}}.
If $R$ belongs to the ideal {in the polynomial ring $\mathbb C[\alpha_0,\ldots,\alpha_n]$, generated by} $Q_{\alpha_0}, Q_{\alpha_1}, ..., Q_{\alpha_n}$, i. e. {if $R$} can be represented {in} the form $R=\sum\limits_{\nu=0}^{n} \lambda_{\nu} Q_{\alpha_{\nu}}$, {where $\lambda_{\nu} $ are polynomials in $\alpha$ and $Q_{\alpha_i}$ stands for $\partial Q/\partial\alpha_i$,} then there exists {a rational} differential $(n-1)$-form $\phi$, such {that } $\gamma - d\phi$ has {a pole} of {order} $k-1$ on the variety $\{ Q(\alpha)=0 \}$.}

\vspace{2mm}\noindent\textit{Proof.} {Assume $R$} belongs to the ideal $(Q_{\alpha_0}, Q_{\alpha_1}, ..., Q_{\alpha_n})$, i. e.  $R=\sum\limits_{\nu=0}^{n} \lambda_{\nu} Q_{\alpha_{\nu}}$ for some polynomials $\lambda_{\nu}$ of {degree} $(k-1)q-n$. Consider the $(n-1)$-form {with pole of order} $k-1$ on the variety $\{ Q = 0\}${:}
\begin{equation} \phi=\frac{1}{k-1}\frac{1}{Q^{k-1}} \sum\limits_{i < j} (-1)^{i+j} [\alpha_i \lambda_j - \alpha_j \lambda_i](... \widehat{d\alpha_i} ... \widehat{d\alpha_j} ...){.} \end{equation}

\noindent We have
\begin{equation}\label{th1exform} d\phi=\left[ \frac{1}{Q^k}\left( \sum\limits_{\nu=0}^{n} \lambda_{\nu} Q_{\alpha_{\nu}} \right) - \frac{1}{k-1}\frac{1}{Q^{k-1}} \sum\limits_{j=0}^{n}\frac{\partial \lambda_j}{\partial \alpha_j}\right] ~\omega. \end{equation}

\noindent So we {obtain}
\begin{equation} \gamma - d\phi = \left[ \frac{1}{k-1}\frac{1}{Q^{k-1}}\sum\limits_{j=0}^{n}\frac{\partial \lambda_j}{\partial \alpha_j} \right] ~\omega. \end{equation}

\hfill$\Box$

\vspace{2mm}  It is necessary to note that under condition $\lambda_{\nu}|_{\alpha_{\nu}=0},~ \nu=0,...,n$ the exact differential form $d \phi$ {does not contribute to} the integral $\int\limits_{\Gamma} \gamma$ by Stokes' theorem, if the hypersurface $\Gamma$ satisfies {the condition} $\partial \Gamma \subset \bigcup\limits_{\nu=0}^{n} \{ \alpha_{\nu}=0 \}$. Besides, the following theorem gives the condition under which  {$R$} belongs to the ideal.

\vspace{2mm}\noindent\textbf{{Macaulay Theorem} \cite{Mac}.~} \textit{Consider $n+1$ homogeneous polynomials $P_0, P_1, ..., P_n$ {on} the projective space $\mathbb{C}\mathbb{P}^n$ of {degrees} $p_0,p_1, ..., p_n$, such that the equations}
$$P_0(\alpha)=0, P_1(\alpha)=0, ..., P_n(\alpha)=0$$
\noindent \textit{have {no solution} in $\mathbb{C}\mathbb{P}^n$. Then all {the} homogeneous polynomials {of degree} greater or equal to  $p_0+p_1+...+p_n-n$ belong to the ideal $(P_0, P_1, ..., P_n)$.}

\vspace{2mm} We {do not give a} proof of this theorem. {The proof consists in an application of the} algorithm of constructing {expansions} of the form $R = \sum\limits_{i=0}^{n} \lambda_i P_i$ for {polynomials $R$ from} the ideal $(P_0, P_1, ..., P_n)$. Applying the {Macaulay} theorem to {the} homogeneous polynomials $Q_{\alpha_0}, Q_{\alpha_1}, ..., Q_{\alpha_n}$ {of degree} $q-1$, we {see} that {any} polynomial $R$ {of degree} $kq-n-1$ belongs to the ideal $(Q_{\alpha_0}, Q_{\alpha_1}, ..., Q_{\alpha_n})$ {provided} $kq-(n+1) \geq (q-2)(n+1) + 1$.

\vspace{5mm}

\section{Partial differential equations for some diagrams}

\vspace{5mm}

Consider {an} arbitrary Feynman diagram and notice that for {any} homogeneous polynomial $P$ {in} $N$ variables {of degree} $p$, we have the following {formula:}
\begin{equation}\label{identity1} P\left( \frac{\partial}{\partial z}\right)F=\int\limits_{\Gamma} P\left( \frac{\partial}{\partial z}\right) \frac{U^{N-\frac{D}{2}(h+1) }}{Q^{N - \frac{D}{2}h}} ~\omega =\frac{(N-\frac{D}{2}h+p-1)!}{(N-\frac{D}{2}h-1)!}\int\limits_{\Gamma} \frac{P(\alpha) U^{N -\frac{D}{2}(h+1) + p} }{Q^{N-\frac{D}{2}h+p}} ~\omega,\end{equation}
\noindent where {$h$, as before, denotes the number of independent loops, and} $\frac{\partial}{\partial z}=\left(\frac{\partial}{\partial z_1}, ..., \frac{\partial}{\partial z_N}\right)$.

\vspace{2mm}\noindent\textbf{{Theorem} 1.~} \textit{The Feynman integral $F(s,z)$ corresponding to {an arbitrary} Feynman diagram is {a solution} of the system of partial differential equations}
\begin{equation*}\left[ Q_{\alpha_i}\left(\frac{\partial}{\partial z}\right) \frac{\partial}{\partial z_i} - U\left(\frac{\partial}{\partial z}\right) - \left( N-\frac{D}{2} (h+1) + q \right) \frac{\partial}{\partial z_i} U_{\alpha_i} \left(\frac{\partial}{\partial z_i}\right) \right] F = 0,~~ i=1,..., N.\end{equation*}

\vspace{2mm}\noindent\textit{Proof.} Consider the differential operator $\mathcal{D}_i=Q_{\alpha_i}\left(\frac{\partial}{\partial z}\right) \frac{\partial}{\partial z_i}$. According to (\ref{identity1}) we have
$$\mathcal{D}_i F = (-1)^{q} \cdot \frac{(N-\frac{D}{2}h+q-1)!}{(N-\frac{D}{2}h-1)!} \int\limits_{\Gamma} \frac{R}{Q^{N-\frac{D}{2}h+q}} ~\omega,$$
\noindent where $R=(-1)^q \alpha_i U^{N-\frac{D}{2}(h+1)+q} Q_{\alpha_i}, \ q=\text{deg}~Q $. It is clear {that} $R$ belongs to the ideal generated {by} $Q_{\alpha_1}, ..., Q_{\alpha_N}$. Moreover, {each polynomial} $\lambda_{\nu}$ {in} the expression $R=\sum\limits_{\nu=0}^{n} \lambda_{\nu} Q_{\alpha_{\nu}}$  {is divisible by} $\alpha_{\nu}$. So after applying the {Griffiths} theorem to the right hand side of (\ref{st4eq1}){, the contribution of} the exact differential form (\ref{th1exform}) {to the integral is zero}. Hence, we have the following
{relation:}
$$\mathcal{D}_i F =  \frac{(N-\frac{D}{2}h+q-2)!}{(N-\frac{D}{2}h-1)!} \int\limits_{\Gamma} \frac{U^{q-1} (U + (N-\frac{D}{2} (h+1) + q ) \alpha_i  U_{\alpha_i})}{Q^{N-\frac{D}{2}h+q-1}} U^{N- \frac{D}{2} (h+1) } ~\omega.$$
\noindent On the other hand, according to (\ref{identity1}) we have
$$\left[U\left(\frac{\partial}{\partial z}\right)+\left(N-\frac{D}{2} (h+1) + q \right) U_{\alpha_i} \left(\frac{\partial}{\partial z_i}\right) \frac{\partial}{\partial z_i} \right]F=$$$$=\frac{(N-\frac{D}{2}h+q-2)!}{(N-\frac{D}{2}h-1)!} \int\limits_{\Gamma} \frac{ U^{q-1} (U + \left( N-\frac{D}{2} (h+1) + q \right) \alpha_i  U_{\alpha_i})}{Q^{N - \frac{D}{2} h + q - 1}} U^{N-\frac{D}{2}(h+1) } ~\omega.$$
\noindent Consequently, comparing the right hands of both previous identities we obtain the statement.

\hfill$\Box$

\vspace{2mm} {Recall that, for a given Feynman diagram, we have introduced a collection $\mathcal{X} \subset 2^{\Theta^{E}}$ of subsets $\chi_1, ..., \chi_r \subset \Theta^{E}$, such that all the invariants $s(\chi)$ are expressed in terms of the invariants $s_i=s(\chi_i)$ and the latter ones are linearly independent.} Now assume that $D \geq n^{E}-1$ and consider the {Feynman} diagrams satisfying the following property{:}

\vspace{2mm}\noindent (P) $\mathcal{X}$ can be chosen in {such} a way that $W_{\chi}=0$ for all $\chi \in 2^{\Theta^E} \setminus \left( \mathcal{X} \cup \{\Theta^{E} - \chi_1, ..., \Theta^{E} - \chi_r \} \right)$.

\vspace{2mm}\noindent In other words, the corresponding polynomial (\ref{denom}) can be written in the form
$$Q = \sum\limits_{i=1}^{r} s_i W_i - U \sum\limits_{i=1}^{N} \alpha_i z_i, \ \ \ \text{where} \ \ \ W_i=W_{\chi_i}.$$
\noindent For example, {the} $h$-loop ladder diagram {(see the figure)} and {the} one-loop $N$-point diagram satisfy this property.
\vspace{4mm}
\begin{center}\includegraphics[scale=0.4]{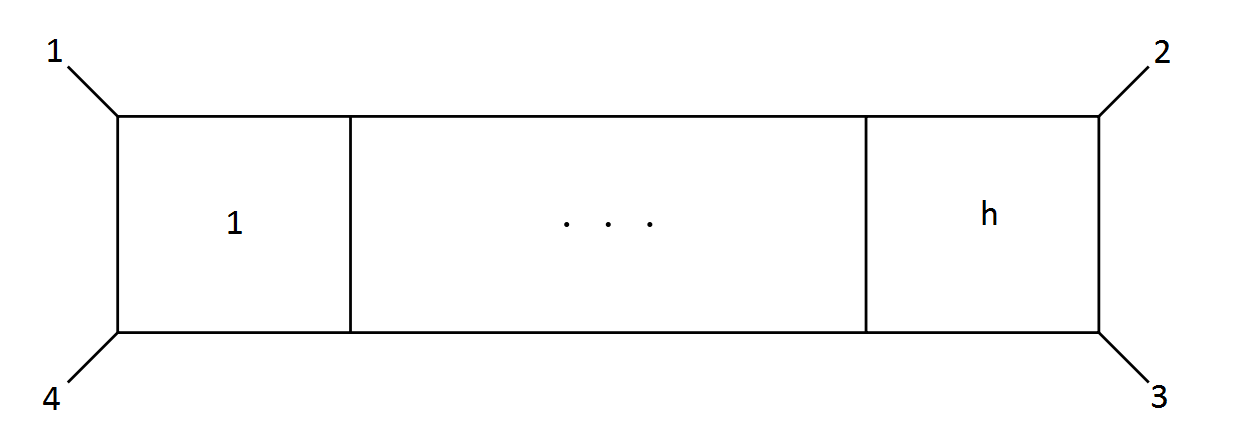}\end{center}
\vspace{1mm}
\noindent As above, notice that for {any} homogeneous polynomial $P$ {in} $r+N$ variables {of} degree $p$ we have the following {equation:}
\begin{equation}\label{identity2} P\left(\frac{\partial}{\partial s}, \frac{\partial}{\partial z}\right)F=(-1)^{p} \cdot \frac{(N-\frac{D}{2}h+p-1)!}{(N-\frac{D}{2}h-1)!}\int\limits_{\Gamma} \frac{P(W,-\alpha U)}{Q^{N-\frac{D}{2}h+p}} U^{N -\frac{D}{2}(h+1) } ~\omega{,}\end{equation}
\noindent where $W=\Big(W_1,...,W_r\Big), \frac{\partial}{\partial s}=\left(\frac{\partial}{\partial s_1}, ..., \frac{\partial}{\partial s_r}\right)$.

\vspace{2mm}\noindent\textbf{{Theorem} 2.~} \textit{Assume that the Feynman diagram satisfies the property (P). Then the corresponding Feynman integral is {a solution} of the system of partial differential equations}
\begin{equation*} \left[ Q_{\alpha_j}\left(\frac{\partial}{\partial z}\right) \frac{\partial}{\partial s_i} + \left( N -\frac{D}{2}(h+1) + q - 1 \right) U_{ \alpha_j}\left(\frac{\partial}{\partial z}\right)\frac{\partial}{\partial s_i} - W_{i, \alpha_j}\left(\frac{\partial}{\partial z}\right) \right] F = 0,  \end{equation*}
\noindent \textit{for all $i, j$ such that the subset $\chi_i$ can be connected with the subset $\Theta^{E}-\chi_i$ by {an} internal line $l_j$.}

\vspace{2mm}\noindent\textit{Proof.} Consider the differential operator $\mathcal{D}_{ij}=Q_{\alpha_j}\left(\frac{\partial}{\partial z}\right) \frac{\partial}{\partial s_i}$. According to (\ref{identity2}) we have
\begin{equation}\label{st4eq1}\mathcal{D}_{ij}F= (-1)^{q} \cdot \frac{(N-\frac{D}{2}h+q-1)!}{(N-\frac{D}{2}h-1)!} \int\limits_{\Gamma} \frac{R}{Q^{N-\frac{D}{2}h+q}} ~\omega,\end{equation}
\noindent where $R=(-1)^{q-1} W_i U^{N-\frac{D}{2}(h+1) + q-1} Q_{\alpha_j}$. It is clear {that $R$} belongs to the ideal generated {by} $Q_{\alpha_1}, ..., Q_{\alpha_N}$. Moreover, due to the property (P){, each} polynomial $W_i$ is {divisible by} $\alpha_j$. So the polynomials $\lambda_{\nu}$ {in} the expression $R=\sum\limits_{\nu=0}^{n} \lambda_{\nu} Q_{\alpha_{\nu}}$ are {divisible by $\alpha_{\nu}$.} Thus after applying the Griffiths' theorem to the right hand side of (\ref{st4eq1}) the exact differential form (\ref{th1exform}) {gives no} contribution {to} the integral.  Hence, we have the following {equality}:
$$\mathcal{D}_{ij}F=(-1) \frac{(N-\frac{D}{2}h+q-2)!}{(N-\frac{D}{2}h-1)!} \int\limits_{\Gamma} \frac{ (U^{q-1} W_{i, \alpha_j} + (N -\frac{D}{2}(h+1) + q - 1) U^{q-2} U_{\alpha_j} W_i )}{Q^{N-\frac{D}{2}h+q-1}} U^{N-\frac{D}{2}(h+1) } ~\omega.$$
\noindent On the other hand, according to (\ref{identity2}) we have
$$\left[ W_{i, \alpha_j}\left(\frac{\partial}{\partial z}\right) - \left( N -\frac{D}{2}(h+1) + q - 1 \right) U_{\alpha_j}\left(\frac{\partial}{\partial z}\right)\frac{\partial}{\partial s_i} \right] F =$$$$= \frac{(N-\frac{D}{2}h+q-2)!}{(N-\frac{D}{2}h-1)!} \int\limits_{\Gamma} \frac{ (U^{q-1} W_{i, \alpha_j} + ( N -\frac{D}{2}(h+1) + q - 1 ) U^{q-2} U_{\alpha_j} W_i )}{Q^{N-\frac{D}{2}h+q-1}} U^{N-\frac{D}{2}(h+1) } ~\omega.$$
\noindent Consequently, comparing the right {hand sides} of both previous identities{,} we obtain the statement.

\hfill$\Box$

\vspace{2mm} Now {we will} present {a} general method {for} deriving {partial} differential equations {of} order $p$ on the example of {a} $h$-loop ladder diagram. This method can be realized by using {a} computer algebra system.

Let $\mathcal{X}=\big\{ \chi_1=\{1\}, ..., \chi_4=\{4\}, \chi_5=\{1, 2\}, \chi_6=\{2, 3\} \big\}$. Then the Feynman integral of {the} $h$-loop ladder diagram has the following form{:}
\begin{equation} F(s,z)=\int\limits_{\Gamma} \frac{U^{h-1} }{Q^{h+1}} ~\omega, \end{equation}
\noindent where $Q=\sum\limits_{i=1}^{6} s_i W_i - U \left( \sum\limits_{i=1}^{3h+1} \alpha_i z_i \right)$, $W_i=W_{\chi_i}$, $s_i=s(\chi_i)$. Consider the differential operators {of} order $p$ and $p-1$ respectively with undetermined coefficients depending on the parameters $s_i, z_j${,}
$$\mathcal{D}=\sum\limits_{|I|+|J|=p} a_{I,J} \partial_{s}^{I} \partial_{z}^{J}, \ \ \ \ \widetilde{\mathcal{D}}=\sum\limits_{|I|+|J|=p-1} b_{I,J} \partial_{s}^{I} \partial_{z}^{J},$$
\noindent where we use the standard multi-index notation. According to (\ref{identity2}) we have
$$\mathcal{D}F=(-1)^{p} \frac{(h+p)!}{h!} \int\limits_{\Gamma} \frac{R}{Q^{h+1+p}} ~\omega, \ \ \ \ \widetilde{\mathcal{D}}F=(-1)^{p-1} \frac{(h+p-1)!}{h!} \int\limits_{\Gamma} \frac{\widetilde{R}}{Q^{h+p}} ~\omega,$$
\noindent where the polynomials $R$ and $\widetilde{R}$ have the following form{:}
$$R= U^{h-1} \left( \sum\limits_{|I|+|J|=p} a_{I,J}^{k} W^{I} \alpha^{J} (-1)^{|J|} U^{|J|} \right), \ \ \ \ \widetilde{R}= U^{h-1} \left( \sum\limits_{|I|+|J|=p-1} b_{I,J}^{k} W^{I} \alpha^{J} (-1)^{|J|} U^{|J|} \right). $$

\noindent Next, consider the homogeneous polynomials $\lambda_{\nu}= \alpha_{\nu} \sum\limits_{|K|=\text{deg}~R - q} \lambda_{\nu, K} \alpha^{K}$ {of} degree $\text{deg}~R - q + 1$ with undetermined coefficients and impose the following conditions {on the polynomials $R, \widetilde{R}, \lambda_{\nu}$: }
$$ R = \sum\limits_{\nu=1}^{3h+1} \lambda_{\nu} Q_{\alpha_{\nu}}, \ \ \ \ \sum\limits_{\nu=1}^{3h+1} \frac{\partial \lambda_{\nu}}{\partial \alpha_{\nu}}= \widetilde{R}.$$
\noindent After collecting coefficients {of} appropriate monomials we obtain {a} system of homogeneous linear equations {with} unknowns $a_{I,J}, b_{I, J}, \lambda_{\nu}^{I}.$ Due to Griffiths theorem each solution of this system corresponds to {a relation}  {of the form} $(\mathcal{D}+\widetilde{\mathcal{D}})F=0$, {which is the wanted  partial differential equation satisfied by the Feynman integral $F$ of a $h$-loop ladder diagram}.

\end{document}